\documentclass[]{article}
\usepackage{amsmath}
\usepackage{amsfonts}
\usepackage{psfrag}
\usepackage{graphicx}
\newcommand{\D}{\,\mathrm{d}}

\newcommand{\onlinecite}[1]{\cite{#1}}

\newcommand{\pdl}[2]{\ensuremath{\partial #1 / \partial #2}}
\newcommand{\pdc}[3]{\ensuremath{\left(\frac{\partial #1}{\partial #2}\right)_{#3}}}
\newcommand{\pdcl}[3]{\ensuremath{\left(\partial #1 / \partial #2\right)_{#3}}}
\newcommand{\jac}[8]{\ \begin{array}{@{\:}r@{}c@{\ }c@{\ }c@{\ }c@{}r@{\:}}\partial(& #1, & #2, & #3, & #4 &)\\\hline\partial(& #5, & #6, & #7, & #8 &)\end{array}\ }
\newcommand{\ie}{\textit{i.e.},~}
\newcommand{\eg}{\textit{e.g.},~}
\newcommand{\vL}{\ensuremath{v_\text{L}}}
\newcommand{\vs}{\ensuremath{v_\text{s}}}
\newcommand{\vsb}{\ensuremath{\mathbf{v}_\text{s}}}
\newcommand{\vnb}{\ensuremath{\mathbf{v}_\text{n}}}
\newcommand{\vsbd}{\ensuremath{\dot{\mathbf{v}}_\text{s}}}
\newcommand{\wc}{\ensuremath{w_\text{c}}}
\newcommand{\HH}{\ensuremath{\mathfrak{H}}}
\renewcommand{\j}{\ensuremath{\mathbf{j}}}

\author{A.F. Andreev and
L.A. Melnikovsky
}
\date{}
\title{Thermodynamics of Superfluidity}
\begin{document}
\maketitle
%\address{
%P.L. Kapitza Institute for Physical Problems\\
%Russian Academy of Sciences, 117334 Moscow, Russia
%}
\psfragscanon
\begin{abstract}
New, superfluid specific additive integral of motion is found. This
facilitates investigation of general thermodynamic equilibrium conditions
for superfluid. The analysis is performed in an extended space of
thermodynamic variables containing (along with the usual thermodynamic
coordinates such as pressure and temperature) superfluid velocity and
momentum density. The equilibrium stability conditions lead to
thermodynamic inequalities which replace the Landau superfluidity
criterion at finite temperatures.

\end{abstract}
%\noindent \textbf{PACS}{: 67.40.Bz, 67.40.Pm, 68.35.Ct, 61.72.Ji}
\section{INTRODUCTION}
Existence of two independent macroscopic motions is a specific feature
of superfluids. Equilibrium state is, thus, characterized by two
velocities of these motions. In general the equilibrium thermodynamic
state is unambiguously determined by the values of all additive integrals
of motion. Integral of motion that corresponds to one of the velocities
(the ``normal component velocity''), like it is in usual hydrodynamics,
is momentum. 

Appearance of the superfluid velocity as a thermodynamic variable
implies the existence of an extra additive integral of motion. Below we
%elucidate the nature of this integral. We 
explicitly find the integral
and this allows us to investigate general thermodynamic stability
conditions and, in particular, the problem of critical velocities.

Usually in experiments the vortices destroy superfluidity at Feynman
critical velocity which is far below the Landau critical velocity. This
is why the superfluid hydrodynamics equations can be expanded in powers
of low velocities  and one safely uses the first nontrivial terms of
this expansion.

Nevertheless, there is a number of experiments\cite{exp1,exp2} where
this is not enough. In some of them the superfluid flow is studied in
small orifices. It has been shown that in these circumstances the
maximum velocity is inversely proportional to the orifice width and may
reach the order of the Landau critical velocity if the aperture is small
enough. Other experiments are performed at temperatures extremely close
to the superfluid transition. In this region, the linear approximation
fails at low velocities starting from those much smaller than the
Feynman critical velocity (which does not depend on the temperature). %
This means that all thermodynamic quantities of the superfluid become
nontrivial functions of the not small superfluid velocity (\ie they
depend not only on the usual thermodynamic coordinates such as pressure
and temperature). The only assumption one can make (and we do it) is
that the fluid at rest is isotropic. This quite general statement of the
problem is used in the paper (and in our earlier letter\cite{jetplet});
in the next section we find the complete set of thermodynamic
inequalities in this light, \ie the conditions imposed on thermodynamic
functions for the superfluid to remain stable. 

Later we employ the Landau phonon-roton model to calculate the highest
velocity compatible with obtained thermodynamic inequalities and show
that it can be interpreted as a critical velocity. This thermodynamic
scenario supposedly explains the superfluidity breakdown in small
orifices. In the vicinity of the $\lambda$-point we use scaling
speculations to find critical behavior of the critical velocity.

\section{STABILITY}
Thermodynamic equilibrium of an isolated system is a state with maximum
entropy at given values of all its additive integrals of motion.

Superfluid system is unusual in the sense it has a continuous gauge
symmetry spontaneously broken, such system possesses a specific quantum
order parameter, ``wave function of the BE condensate''. This is why in
superfluid there are two independent velocities of normal and superfluid
motion. The superfluid velocity is (up to a factor) a gradient of the
wave function phase $\vsb=\nabla \phi$. The same is true for the time
derivative $\vsbd=\nabla \dot{\phi}$. This is a local conservation
law of all three components of the vector ${\mathbf{V}}_\text{s}=\int
{\mathbf{v}}_\text{s} \D \mathbf{r}$. In other words, $\vsb$ is the
density of a new additive integral of motion (along with mass density or
momentum density). This integral is specific to superfluid.

Instead of the entropy maximality,  it is usually  convenient\cite{LL5}
to use another requirement, equivalent to the first one, namely the
energy minimality under constant entropy and additive integrals of
motion. This is the requirement of thermodynamic stability and may be
expressed in form of {\em thermodynamic inequalities}.

Total energy of the superfluid $E_\text{tot}$ is an integral of the energy density $E$ over the
entire volume $E_\text{tot}=\int E \D \mathbf{r}$.
The energy density can be obtained via a Galilean
transformation\cite{khalat}
%is given by equation
\begin{equation}\label{Edensity}
E=\frac{\rho \vs^2}{2}+\vsb\j_0+E_0.
\end{equation}
Here, \vsb\ is the superfluid velocity, $\rho$ is the mass density and
subscript ``0'' denotes quantities measured in the frame of reference of
the superfluid component (that is the frame, where the superfluid
velocity is zero). Namely, $E_0$ and $\j_0$ are the energy density and
the momentum density with respect to the superfluid component. The
former is a function of $\rho$, $\j_0$, and the entropy density $S$. Its
differential can be written as
\begin{equation}\label{E0density}
\D E_0= T \D S + \mu\D \rho +  \mathbf{w} \D \j_0,
\end{equation}
where Lagrange multipliers $T$, $\mu$, and $\mathbf{w}$ are the
temperature, the chemical potential, and the so-called relative velocity
of normal and superfluid components.

The liquid is isotropic and, consequently, the velocity $\mathbf{w}$
and the momentum density $\j_0$ are parallel to each other:
\begin{equation*}
\j_0=j_0(T,\rho,w)\frac{\mathbf{w}}{w}.
\end{equation*}
This leads to a useful identity for the partial derivatives of $\j_0$ with
respect to $\mathbf{w}$:
\begin{equation} \label{djw}
\pdc{j_0^k}{w^l}{T,\rho}=\frac{w^k w^l}{w^2} \pdc{j_0}{w}{T,\rho}
+\left(\frac{\delta^{kl}}{w} - \frac{w^k w^l}{w^3}\right)j_0.
\end{equation}

Further transforming \eqref{Edensity}, we can write
$\D E$ with the help of \eqref{E0density} in
the form
\begin{equation}
\D E= T\D S +
\left(\mu+\frac{\vs^2}{2}-\vsb\vnb\right)
\D\rho +(\j-\rho\vnb)\D\vsb +
\vnb\D \j,
\end{equation}
where
$\j=\rho\vsb + \j_0$
is the total momentum density and $\vnb=\vsb+\mathbf{w}$ is
the normal velocity.

Stability implies that each ``allowed'' fluctuation increases the total
energy of the system $E_\text{tot}$. Allowed are the fluctuations
leaving conserved quantities unchanged. This means that the minimality
of $E_\text{tot}$ must be investigated under fixed entropy and all
additive integrals of motion: mass, momentum, and superfluid velocity.

Consider a macroscopic fluctuation of all the variables $\delta S$, $\delta \rho$,
$\delta \vsb$, and $\delta \j$. They are conserved and this ensures that
the first variation of the total energy for a uniform system is identically zero
\begin{multline*}
\delta E_\text{tot}=\int \biggl\{
                                           \pdc{E}{S}{\rho,\vsb,\j}\delta S
                                          +\pdc{E}{\rho}{S,\vsb,\j}\delta \rho+\\
                                           \pdc{E}{\vsb}{S,\rho,\j}\delta \vsb
                                          +\pdc{E}{\j}{S,\rho,\vsb}\delta \j
                                       \biggr\} \D \mathbf{r} \equiv 0.
\end{multline*}
The matrix of the second variation quadratic
form is a Jacobian matrix $8\times 8$
\begin{equation*}
\HH=
\left\|\jac{T}{\vnb}{\ \mu+\vs^2/2-\vsb\vnb}{\ \j-\rho\vnb}
      {S}{\j}{\rho}{\vsb}\right\|.
\end{equation*}
It is positive definite
if all principal minors in the top-left corner
$M_1, M_2,\dots M_8$  are
positive.
We recursively test these minors:
%%%%%%%%%%%%%%%%%%%%%%%%%%%%%%%%%%%
\begin{itemize}
\item The first positivity condition
\begin{multline*}
M_1=\jac{T}{\j}{\rho}{\vsb}{S}{\j}{\rho}{\vsb}=
\jac{T}{\j}{\rho}{\vsb}{T}{\vnb}{\rho}{\vsb}
\jac{T}{\vnb}{\rho}{\vsb}{S}{\j}{\rho}{\vsb}=\\
\pdc{j_0}{w}{T,\rho}
\left( 
  \pdc{S}{T}{\rho,w}\pdc{j_0}{w}{T,\rho} - \pdc{j_0}{T}{\rho,w}^2
\right)^{-1}
%=\left(\pdc{S}{T}{\rho,w} - \pdc{j_0}{w}{T,\rho}^{-1} \pdc{j_0}{T}{\rho,w}^2\right)^{-1}
>0
\end{multline*}
corresponds to the usual requirement of the heat capacity positivity. It is shown below
that $\pdcl{j_0}{w}{T,\rho}>0$, hence the last inequality eventually becomes
\begin{equation}\label{M1}
\pdc{S}{T}{\rho,w}\pdc{j_0}{w}{T,\rho} - \pdc{j_0}{T}{\rho,w}^2>0.
\end{equation}

\item 
Positivity of the next group of minors is easily verified with the following transformation
\begin{equation}\label{Q1}
\HH'=\left\|\jac{T}{\vnb}{\rho}{\vsb}{S}{\j}{\rho}{\vsb}\right\|=
\left\|\jac{T}{\j}{\rho}{\vsb}{S}{\j}{\rho}{\vsb}\right\|
\left\|\jac{T}{\vnb}{\rho}{\vsb}{T}{\j}{\rho}{\vsb}\right\|.
\end{equation}
Positivity of the minors $M_2, M_3, M_4$ is guaranteed
if all minors of the second term in \eqref{Q1} are positive
\begin{multline*}
\left\|\pdc{\j}{\vnb}{T,\rho,\vsb}\right\|^{-1}=
\left\|\pdc{\j_0}{\mathbf{w}}{T,\rho}\right\|^{-1}=\\
\begin{Vmatrix}
\pdcl{j_0}{w}{T,\rho} &0      &0\\
0                     &j_0/w  &0\\
0                     &0      &j_0/w
\end{Vmatrix}^{-1}.
\end{multline*}
Here we used \eqref{djw} and chose the direction of the $\mathbf{w}$
vector as the first coordinate. Our collection is therefore augmented by
two more inequalities
\begin{equation}\label{M2a}
\j_0\mathbf{w}\geq 0,
\end{equation}
\begin{equation}\label{M2b}
\pdc{j_0}{w}{T,\rho}>0.
\end{equation}

\item The same transformation applied to the biggest minors yields:
\begin{equation*}
\HH=
\HH'
\left\|\jac{T}{\vnb}{\mu+\vs^2/2-\vsb\vnb}{\j-\rho\vnb}{T}{\vnb}{\rho}{\vsb}\right\|=
\HH' \HH''.
\end{equation*}
Again, the minors $M_5,M_6, M_7,$ and $M_8$ are positive if all
nontrivial principal minors of $\HH''$ are positive.
We use the thermodynamic identity to relate the chemical potential~$\mu$
and the conventional pressure~$p$
\begin{equation*}
\D\mu=\frac{\D p}{\rho}-\frac{S}{\rho}\D T -
\frac{\j_0}{\rho}\D\mathbf{w},
\end{equation*}
This gives
\begin{equation*}
\pdc{\left(\mu+\vs^2/2-\vsb\vnb\right)}{\rho}{T,\vnb,\vsb}=
\pdc{\mu}{\rho}{T,\mathbf{w}}=\frac{1}{\rho}\pdc{p}{\rho}{T,w}.
\end{equation*}
The following is an explicit representation of $\HH''$ sub-matrix corresponding
to a four-dimensional sub-space
$\rho,\vs^x,\vs^y,\vs^z$; as before we let the $x$-axis run along $\mathbf{w}$ direction.
Using \eqref{djw} we obtain
\begin{equation*}
\begin{Vmatrix}
\pdcl{p}{\rho}{T,w}/\rho &\pdcl{j_0}{\rho}{T,w}-w               &0 &0\\
\pdcl{j_0}{\rho}{T,w}-w  &\rho-\pdcl{j_0}{w}{T,\rho}    &0 &0\\
0 &0 &\rho-j_0/w &0\\
0 &0 &0          &\rho-j_0/w
\end{Vmatrix}
.
\end{equation*}

Appropriate inequalities are:
\begin{equation}\label{M3}
\pdc{p}{\rho}{T,w}>0,
\end{equation}
which is literally a generalized (to a non-zero inter-component velocity $w$)
positive compressibility requirement,
\begin{equation}\label{M4a}
j_0<w\rho,
\end{equation}
\begin{equation}\label{M4b}
\pdc{p}{\rho}{T,w}
\left(\rho-\pdc{j_0}{w}{T,\rho}\right)
-\rho\left(\pdc{j_0}{\rho}{T,w}-w\right)^2>0.
\end{equation}
\end{itemize}

Inequalities \eqref{M1}, \eqref{M2a}, \eqref{M2b},
\eqref{M3}, \eqref{M4a}, and \eqref{M4b} are 
sufficient conditions for the thermodynamic stability.

\section{DISCUSSION}
In a ``stopped-normal-component'' arrangement, the mass flux $\mathbf{f}$ with respect to the normal
component may become more convenient than $\j_0$---the mass flux relative to the superfluid one.
The obvious relation between them $f=\rho w - j_0$ leads to the following reformulation of the inequalities:

\begin{equation}\label{F1}
\mathbf{fw}<0, \quad f<w\rho,
\end{equation}
\begin{equation}\label{F2}
0<\pdc{f}{w}{\rho,T}<\rho,
\end{equation}
\begin{equation}\label{F3}
\pdc{S}{T}{\rho,w}\left(\rho-\pdc{f}{w}{T,\rho}\right) > \pdc{f}{T}{\rho,w}^2,
\end{equation}
\begin{equation}\label{F4}
\pdc{p}{\rho}{T,w}\pdc{f}{w}{\rho,T}>\rho\pdc{f}{\rho}{w,T}^2.
\end{equation}

As a simple application of the derived inequalities, consider them at $w=0$.
From \eqref{F1}, \eqref{F2},  \eqref{F3}, and  \eqref{F4} we get
\begin{equation*}
\pdc{S}{T}{\rho,w}>0, \quad 
\pdc{p}{\rho}{T,w}>0,
\end{equation*}
\begin{equation*}
\rho>\pdc{j_0}{w}{T,\rho}>0.
\end{equation*}
Using conventional notation, last inequality reads
\begin{equation}\label{lambdapoint}
\rho_\text{s}>0, \quad \rho_\text{n}>0.
\end{equation}

\section{PHONON-ROTON MODEL}
Here, we provide a practical example of the stability criteria for
real superfluid ${}^4\mathrm{He}$.
To calculate derivatives involved in the inequalities one must use
the microscopic approach. Simple and clear Landau phonon-roton model
works pretty well
over wide temperature and velocity ranges. We use this model
to calculate the contribution of these quasiparticles to the ``modified'' free
energy in the frame of reference of the superfluid component:
\begin{equation*}
\widetilde{\cal F}_{0}= {\cal F}_{0}-\mathbf{w j}_0,\quad
\D\widetilde{\cal F}_{0}=-S\D T - \j_0\D \mathbf{w}.
\end{equation*}
This potential is obtained from the excitation spectrum with the
conventional formula
\begin{equation*}
\widetilde{\cal F}_{0}=
T\int\ln\left(1-\exp\left(\frac{\mathbf{pw}-\epsilon(p)}{T}\right)\right)\frac{\D\mathbf{p}}{(2\pi\hbar)^3}.
\end{equation*}
We denoted the excitation energy $\epsilon(p)$, which is given for two branches by the expressions
\begin{equation*}
\epsilon_\text{ph}(p)=cp,
\quad
\epsilon_\text{r}(p)=\Delta+{(p-p_0)^2}/(2m).
\end{equation*}
Subscripts distinguish the quantities related to phonons and rotons, $c$
is the sound velocity, $\Delta$ is the roton energy gap, $m$ is the
effective mass, and $p_0$ is the momentum at the roton minimum  (we take
numerical data from Refs.~\onlinecite{don76},~\onlinecite{don81}:
$\rho=0.145\,\text{g}/\text{cm}^3$,
$\Delta=8.7\,\text{K}$,
$m=0.16 m_\text{He}$,
%$p_0=3.673\cdot 10^{8}\,\text{g}^{-1/3}\rho^{1/3}\hbar$,
$p_0\rho^{-1/3}\hbar^{-1}=3.673\cdot 10^{8}\,\text{g}^{-1/3}$,
$c=238\,\text{m}/\text{s}$,
$\pdl{\Delta}{\rho}= -0.47\cdot 10^{-14}\,\text{cm}^5\text{s}^{-2}$,
$\pdl{m}{\rho}= -0.45\cdot 10^{-23}\,\text{cm}^3$,
$\pdl{c}{\rho}= 467\cdot 10^{3}\,\text{cm}^4 \text{s}^{-1}\text{g}^{-1}$).
A small dimensionless parameter $m\Delta / p_0^2 \sim 0.03 \ll 1$
ensures, \eg that the Landau critical velocity is determined by
$\vL=\Delta/p_0$.

When integrated, these dispersion laws give the following contribution to the free energy:
\begin{equation*}
\widetilde{\cal F}_{0}=
-\frac{T^4\pi^2}{90\hbar^3c^3}\left(1-\frac{w^2}{c^2}\right)^{-2}
-\frac{T^{5/2}m^{1/2}}{2^{1/2}\pi^{3/2}\hbar^3}
 %\left( m\cosh\frac{wp_0}{T}+
 \frac{p_0}{w}\sinh\frac{wp_0}{T}
 %\right)
 \exp\left(
 %\frac{mw^2}{2T}
 -\frac{\Delta}{T}\right)
 .
\end{equation*}
One can obtain all thermodynamic variables by differentiating this
potential. Namely
\begin{equation*}
S=-\pdcl{\widetilde{\cal F}_0}{T}{w,\rho},\quad j_0=-\pdcl{\widetilde{\cal F}_0}{w}{T,\rho}.
\end{equation*}
We neglect the quasiparticle contribution to the pressure derivative.

\begin{figure}[ht]
\begin{center}
\includegraphics[scale=0.5]{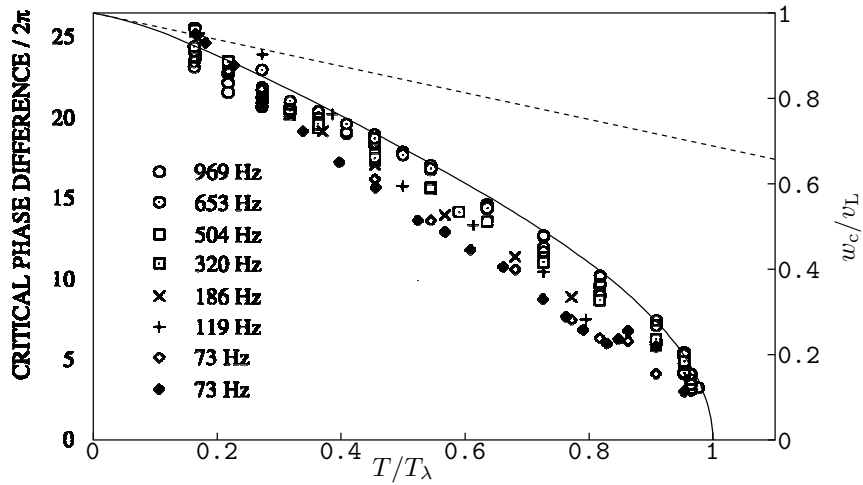}
\end{center}
\caption{Critical velocity \wc\ versus temperature $T$. Dashed line
corresponds to the equation $T=\Delta-p_0 w$. Note that the condition
$T<\Delta-p_0 w$ holds true over entire stability domain. The
``stability'' critical velocity \wc\ coincides with the Landau critical
velocity \vL\ at zero temperature and vanishes completely (see
\eqref{lambdapoint}) at the critical temperature $T_\lambda$. The curve
is superimposed on the experimental graph of the critical phase
difference over a $2\ \mu\text{m} \times 2\ \mu\text{m}$ aperture in a thin
foil at various frequencies.\cite{zimmer}}
\label{Wcrit}
\end{figure}

\begin{figure}[h]
\begin{center}
\includegraphics[scale=0.6]{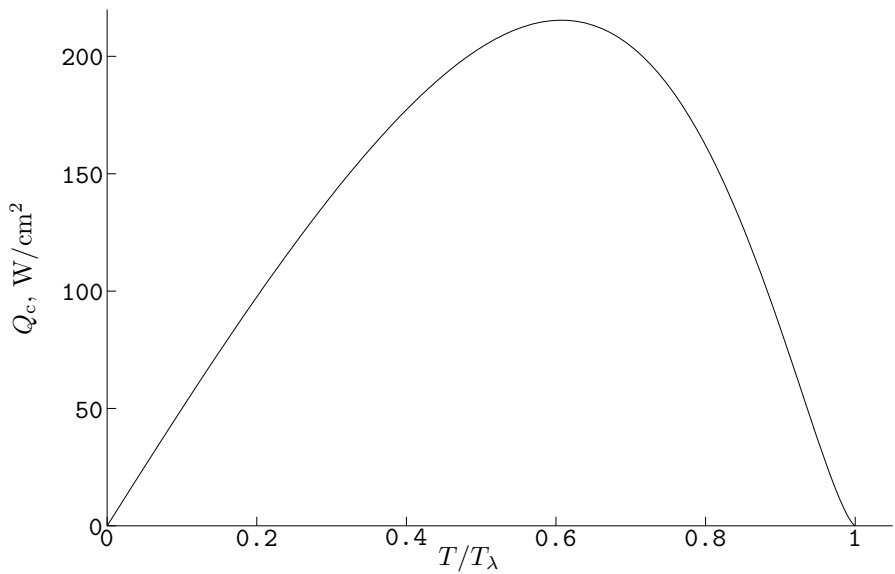}
\end{center}
\caption{Critical energy flux (at zero mass flux)$Q_c$ versus temperature $T$.}
\label{Qcrit}
\end{figure}

At zero velocity the inequalities in Eq. \eqref{lambdapoint} fail at the
$\lambda$-point. When the velocity is higher, the inequality \eqref{M4b}
is the first to become invalid. At zero temperature, the critical (the
highest possible without stability violation) velocity tends to the
Landau critical velocity \vL\footnote{Note that for systems where all
quasiparticles can be described hydrodynamically (in other words,
systems without roton branch) inequality \eqref{M4b} at zero temperature
includes a condition $\pdcl{p}{\rho}{T,w}-w^2>0$, \ie $w<c$.}. Entire
stability region is plotted in Fig.~\ref{Wcrit} and Fig.~\ref{Qcrit}.
The liquid is unstable above the curves.

In Fig.~\ref{Wcrit} the critical velocity is plotted with the solid
line. Scaled experimental data\cite{zimmer} are presented in the same
graph. High frequency points follow the shape of the theoretical curve,
but at low frequencies sufficient discrepancy is observed. This might be
explained by the long time required for vortex nucleation. At higher
frequencies the thermodynamic stability limit is observed while lower
frequency critical velocity is cut off by vortices. Whatever feasible
explanation this might be, the following warning must be taken into
account when experimental data on critical phase difference and our
predictions are compared. Assumption that the critical velocity is
determined by the stability limit implies that the hydrodynamic
equations inside orifice are essentially nonlinear. Particularly, one can
not consider the superfluid component as an incompressible fraction of
the liquid. In other words the phase shift across the length of the
orifice is {\em not} proportional to the maximum velocity attained in the
orifice.

In Fig.~\ref{Qcrit} the critical energy flux is plotted. In typical
arrangement the energy is carried by the superfluid counterflow, where
normal and superfluid velocities are directed oppositely to each other
to keep zero net mass flux. In zero mass flux frame the condition $j=0$
yields
\begin{equation}
\label{eflux}
Q=TS\left(w-\frac{j_0}{\rho}\right) +
\left(w-\frac{j_0}{\rho}\right)^2 j_0
\end{equation}
for the energy flux.

\section{CRITICAL BEHAVIOR}
To find the behavior of the critical velocity near the $\lambda$-point, 
scaling arguments should be used. We assume that the low velocity
expansion of the irregular part of thermodynamic potential near
$T_\lambda$ starts with
\begin{equation}
\label{cexpansion}
A \tau^{2-\alpha} + \rho_s\frac{\vs^2}{2} - B \tau^x \vs^4,
\end{equation}
where $A$ and $B$ are constants, $\tau=(T_\lambda-T)/T_\lambda$,
$\alpha$ is the critical index of the specific heat $C_p \propto
\tau^{-\alpha}$,  the ``superfluid density'' $\rho_s$ follows Josephson
scaling relation\cite{joseph} $\rho_s \propto \tau^{(2-\alpha)/3}$, and
the index $x$ will be determined below. According to the hypothesis of
scale invariance, all terms in \eqref{cexpansion} are of the same order
in the fluctuation region. This yields $\tau^x \propto \rho_s^2\tau^{\alpha-2}$
and $x=(\alpha-2)/3$. On the other hand, the stability is destroyed
by the velocity at which the linear approximation fails, \ie 
\begin{equation}
w_c \propto \sqrt{\rho_s / \tau^x} \propto \tau^{(2-\alpha)/3}.
\end{equation}
Current value for the specific heat index\cite{alpha} $\alpha=-0.0127
\pm 0.0003$ gives 0.671 exponent for the critical velocity

Extensive data near the  $\lambda$-point are available in the critical
energy flux experiments.\cite{exp2} In the vicinity of the transition,
Eq. \eqref{eflux} is reduced to $Q \approx TS(w-j_0/\rho)$ and the
critical flux is determined by
\begin{equation}
Q_c \propto \rho_s w_c \propto \tau^{(4-2\alpha)/3} \approx \tau^{1.342},
\end{equation}
which is in moderate agreement with experimental value of the exponent
$1.23 \pm 0.02$.

\section{CONCLUSION}
Experimentally, the superfluidity breakdown is believed to have the
following nature.\cite{exp1} Until the aperture size is too small, the
critical velocity does not depend on the temperature and increases as
the size decreases. This is the very behavior that is specific to the
Feynman vortex-related critical velocity.

When the orifice width is narrow enough, or the temperature is close
enough to $T_\lambda$ the breakdown scenario and its features change. The
critical velocity does not depend on the aperture any more but decreases
when the temperature increases. This behavior is commonly
associated\cite{exp1} with the Iordanski-Langer-Fisher
mechanism.\cite{JLF} Nevertheless, this association lacks numerical
comparison because no reliable information about the actual orifice
shape is available.

On the other hand experimentally observed behavior of the critical
velocity can be attributed to the suggested stability criterion. In
other words, we provide an alternative explanation of experimental
results based on an assumption that in narrow orifices the thermodynamic
limit of \wc\ is reached.

We should also note that our approach to the critical velocity as a
stability limit is similar to that used by Kramer.\cite{kramer}
Actually, the inequality he employed is not a thermodynamic one, but
numerical results for the critical velocity he obtained using the
phonon-roton model do not deviate much from those plotted in
Fig.~\ref{Wcrit}.

\section*{ACKNOWLEDGMENTS}
Opportunity to write the paper for the JLTP issue dedicated to the
memory of Olli Lounasmaa, the person contributed so much to the physics
of superfluidity and to the field of low temperatures in general, is a
great honor to us. It is also very important for us that Olli was one
of the founders of the close scientific collaboration between Finland
and Russia which turned out to be so important for the world science.

It is a pleasure for us to thank I.A. Fomin for useful discussions. One
of us (LM) would also like to thank University of New Mexico for its
hospitality during writing this paper. This work was supported in parts
by INTAS grant 01-686, CRDF grant RP1-2411-MO-02, RFBR grant
03-02-16401, and RF president program.

\end {document}